\documentstyle[twoside,fleqn,espcrc2,fps]{article}

\newcommand{\unit}{1 \!\! 1}

\newcommand{\AmS}{{\protect\the\textfont2
  A\kern-.1667em\lower.5ex\hbox{M}\kern-.125emS}}

\title{The Perfect Quark-Gluon Vertex Function
        \thanks{Based on a poster presented by K. Orginos at LAT97.}}

\author{K. Orginos \address{Dept. of Physics, Brown University,
        Providence, RI 02912, USA},
        W. Bietenholz \address{HLRZ c/o Forschungszentrum J\"{u}lich,
        52425 J\"{u}lich, Germany},
        R. Brower \address{Dept. of Physics, Boston University,
        Boston, MA 02215, USA},
        S. Chandrasekharan \address{CTP-LNS, MIT, Cambridge, MA 02139, USA}
        and 
        U.-J. Wiese${\rm ^{d}}$}

\begin{document}

\begin{abstract}
We evaluate a perfect quark-gluon vertex function for QCD
in coordinate space and truncate it to a short range.
We present preliminary results for the charmonium spectrum
using this quasi-perfect action.
\end{abstract}

\maketitle

Approximately perfect lattice actions have a potential
to suppress lattice artifacts much more than the $O(a)$
improvement, which is very fashionable at this conference.
However, it is still unclear how well this more sophisticated
improvement program can be applied to QCD.
Here we discuss our recent progress in the construction of a
quasi-perfect action for QCD, and show preliminary results
of its application to heavy quarks. \\

\begin{figure}[hbt]
\def\fpsangle{0}
\epsfxsize=70mm
\fpsbox{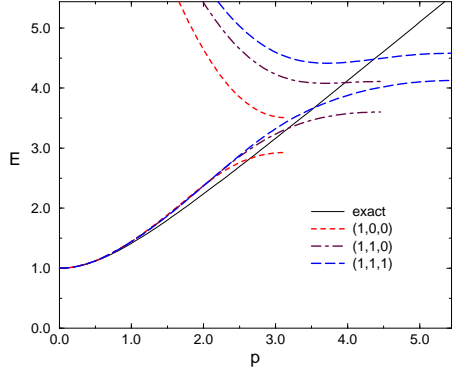}
\vspace{-10mm}
\caption{The dispersion relation of the hypercube fermion
at $m=1$ in various directions.}
\label{dispfig}
\vspace{-8mm}
\end{figure}

For free fermions we have derived extremely local perfect lattice
actions \cite{QuaGlu} of the form,
\begin{displaymath}
S[\bar \Psi , \Psi ] = \sum_{x,y} \bar \Psi_{x}
[ \gamma_{\mu} \rho_{\mu}(y-x)+\lambda (y-x)] \Psi_{y}.
\end{displaymath}
The couplings decay exponentially (in $d>1$). We truncate them
to a unit hypercube by means of 3-periodic boundary conditions.
The resulting ``hypercube fermion'' has still excellent
spectral and thermodynamic properties \cite{LAT96}.
%
%
As an example, we give the couplings for $m=1$ in Table \ref{coptab}
and show the dispersion relation in Fig.
\ref{dispfig}. It is very close to the continuum dispersion
and approximates rotational symmetry very well.

\begin{table}
\begin{center}
\begin{tabular}{|c|c|c|}
\hline
$ y-x $ & $\rho_{1}(y-x)$ & $\lambda (y-x)$ \\
\hline
\hline
(0000) & ~0             & ~1.26885069540 \\
\hline 
(1000) & ~0.05457967484 & -0.03008271460 \\
\hline
(1100) & -0.01101007028 & -0.01082956270 \\
\hline
(1110) & -0.00325481234 & -0.00471575763 \\
\hline
(1111) & -0.00120632489 & -0.00221240767 \\
\hline
\end{tabular}
\end{center}
\caption{The ``hypercube fermion'' couplings at $m=1$.}
\label{coptab}
\vspace{-10mm}
\end{table}

%
%
%
%

Expansion to $O(g A_{\mu})$ for full QCD yields
\begin{eqnarray*}
S[\bar \Psi ,\Psi , A_{\mu}] &=& S[\bar \Psi , \Psi ]
+ S[A_{\mu}] + g V[\bar \Psi ,\Psi ,A_{\mu}] \\
V[\bar \Psi , \Psi ,A_{\mu}] &=& \frac{1}{(2\pi )^{2d}}
\int_{B^{2}} dp dq  \bar \Psi^{i}(-p) 
V_{\mu}(p,q) \\ && A_{\mu}^{a}(p-q)
\lambda_{ij}^{a} \Psi^{j}(q) .
\end{eqnarray*}
The perfect action for free gluons, $S[A_{\mu}]$, has also been 
discussed in \cite{QuaGlu}. Moreover,
we found an explicit but complicated expression
for the {\em perfect quark-gluon vertex function 
$V_{\mu}(p,q)$}.

Numerically the vertex function can be evaluated and transformed
to c-space, where it is re-presented by a set of link couplings,
which depend on the fermion positions.

We truncate the $O(gA_{\mu})$ perfect action and parameterize it
in a gauge invariant form:

1) Take the mean value 
of the link couplings over short
lattice paths connecting $\bar \Psi$ and $\Psi$.

2) Re-scale the path and plaquette couplings
such that their sum amounts to $\lambda (r)$ for the
scalar terms ($\propto \unit$), 
$\rho_{\mu}(r)$ for the vector terms
($\propto \gamma_{\mu}$),
$s(m) = (m/\hat m )^{2} [1/m - 1/\hat m ]$, 
$ \hat m \doteq e^{m}-1$
for the plaquette couplings ($\propto \sigma_{\mu \nu}$) \cite{LAT96}
and a value 
$s_{1}(m)$ -- obtained from a low 
$\vec p$ expansion -- for the terms $\propto \gamma_{\mu}\gamma_{\nu}
\gamma_{\rho}$. 

For our first experiments we impose a tough truncation and arrive
at the following parameterization of $V_{\mu}(x,y)$:

\begin{center}
\begin{tabular}{|c|c|c|c|c|}
\hline
& & $m=0$ & $m=1$ & $m=2$ \\
\hline
\hline
$\unit$ & $L_{1}$ & -0.01315 & -0.00528 & -0.00184 \\
\hline
& $L_{3}^{s}$ & -0.00793 & -0.00413 & -0.00169 \\
\hline
& $L_{2}$ & -0.01502 & -0.00541 & -0.00163 \\
\hline
& $L_{3}$ & -0.00266 & -0.00079 & -0.00019 \\
\hline
& $L_{4}$ & -0.00035 & -0.00009 & -0.00002 \\
\hline
\hline
$\gamma_{\mu}$ & 
$L_{1}$ & ~0.02982 & ~0.01086 & ~0.00338 \\
\hline
& $L_{3}^{s}$ & ~0.01784 & ~0.00729 & ~0.00253 \\
\hline
& $L_{2}$ & ~0.01604 & ~0.00551 & ~0.00158 \\
\hline
& $L_{3}$ & ~0.00184 & ~0.00054 & ~0.00013 \\
\hline
& $L_{4}$ & ~0.00020 & ~0.00005 & ~0.00001 \\
\hline
\hline
$\sigma_{\mu \nu}$ & $c_{0}$ & ~0.07283 & ~0.02207 & ~0.00556 \\
\hline
& $c_{1}$ & ~0.05217 & ~0.01333 & ~0.00285 \\
\hline
\hline
$\gamma_{\mu} \gamma_{\nu} \gamma_{\rho}$ & $s_{1}$ 
& ~0.05457 & ~0.01335 & ~0.00268 \\
\hline
\end{tabular}
\end{center}
For the scalar term, $L_{1}$ is a path consisting of one link,
$L_{3}^{s}$ is a staple and $L_{2},~L_{3},~L_{4}$ are the shortest
paths connecting $\bar \Psi , ~ \Psi$ separated by a 2d, 3d, 4d
diagonal in the unit hypercube, respectively.
The same holds for the vector term, if the fermions 
are separated in the $\mu$ direction.
The standard clover-like plaquette coupling is called $c_{0}$,
and $c_{1}$ refers to the case where $\bar \Psi ,~ \Psi$ are
separated by one lattice spacing; it is the coupling to the 
plaquettes attached to their connecting link.
Finally, if the fermions are separated by $\hat \rho$, then
we couple them to the plaquettes in the perpendicular $\mu ,\nu$ planes
(touching the fermions), and to the connecting link, by $s_{1}$.

A low $\vec p$ expansion leads to a Hamiltonian of the form
\vspace*{-1mm}
\begin{displaymath}
H = m_{s} + \frac{1}{2m_{kin}} 
\vec p^{~2} + \frac{1}{2m_{B}}
\vec \Sigma \vec B + \dots
\end{displaymath}
\vspace*{-1mm}
where $\vec B$ is assumed const. in time, and $\Sigma_{k}
= \epsilon_{ijk}\sigma_{ij}/2$, \cite{Fermilab}.
Here $m_{s}$ is the static, $m_{kin}$ the kinetic and $m_{B}$
the magnetic mass. In the continuum they all coincide, and
for the hypercube fermion $m_{s}=m$.

\begin{figure}[hbt]
\vspace{-6mm}
\def\fpsangle{0}
\epsfxsize=70mm
\fpsbox{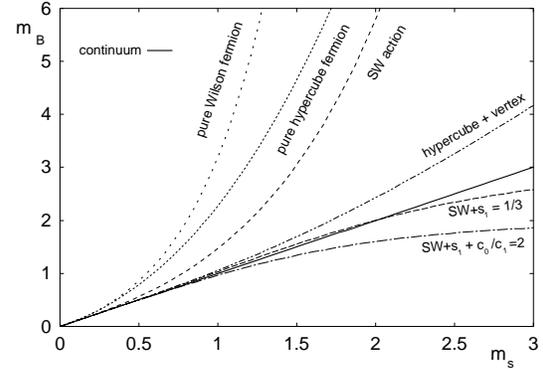}
\vspace{-10mm}
\caption{The magnetic vs. static mass.}
\label{mBfig}
\vspace{-8mm}
\end{figure}

%
We saw before that $m_{kin}(m_{s})$ is drastically improved
for the hypercube fermion, compared to the Wilson fermion 
\cite{LAT96}.
Of particular interest for the hyperfine splitting is $m_{B}(m_{s})$,
see Fig. \ref{mBfig}.

Remarkably, the Wilson fermion can be corrected up to the
quartic order by the parameters we introduced before: \\
pure Wilson fermion: $m_{B} = m_{s} + m_{s}^{2} + \dots$ \\
Sheikholeslami-Wohlert action ($c_{0}=1$): 
$m_{B} = m_{s} + \frac{2}{3}m_{s}^{3} + \dots$ 
(and $m_{B}=m_{kin}$). \\
 " " + $s_{1}=\frac{1}{3}$: 
$m_{B} = m_{s} + \frac{1}{6}m_{s}^{4} + \dots $ \\
split clover term into $c_{0}=\frac{2}{3},\ c_{1}=\frac{1}{3}$: 
$m_{B} = m_{s} - \frac{4}{45} m_{s}^{5} + \dots $ \\

\vspace*{-2mm}
We now present preliminary results on {\em charmonium spectroscopy}
using the fermionic action presented above and the Wilson
gauge action. We simulate quenched at $\beta = 5.6$, the lattice spacing
is 0.24 fm and the size $8^{3}\times 16$. We include 30 lattices
and consider the meson dispersion relation as well as the spectrum
of the low lying states. The latter was evaluated at $m=0$ and 
$m=1$. We then interpolate to the mass of $\eta_{c}$ (2.98 GeV),
which corresponds to $m\simeq 0.9$.

\begin{figure}[hbt]
\vspace{-6mm}
\def\fpsangle{0}
\epsfxsize=60mm
\fpsbox{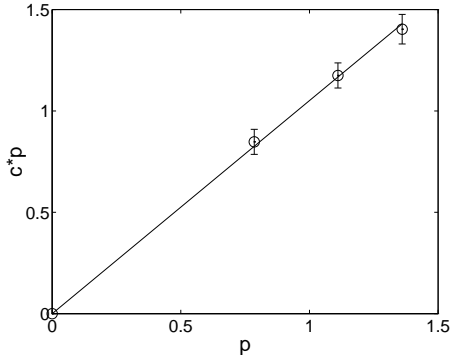}
\vspace{-10mm}
\caption{The mesonic dispersion relation.}
\label{mesdisp}
\vspace{-8mm}
\end{figure}

Regarding the dispersion relation we compute the ``speed of light''
from $cp = \sqrt{E^{2}-M^{2}}$, see Fig. \ref{mesdisp},
and obtain $c=1.04 \pm 0.05$.
The mesonic dispersion is impressively accurate even for large
momenta. 

\begin{figure}[hbt]
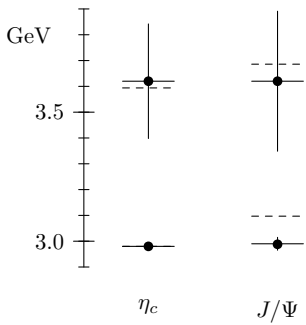

\vspace{-5mm}
\def\fpsangle{0}
\epsfxsize=40mm
\fpsbox{spectrum.3}
\vspace{-10mm}
\caption{The charmonium spectrum. The experimental values are dashed,
and only the ground state of $\eta_{c}$ is fitted.}
\label{chaspec}
\vspace{-8mm}
\end{figure}

In Fig. \ref{chaspec} we show the charmonium spectrum,
including the 1s and 2s states of $\eta_{c}$ and $J/\psi$.
The 2s-1s splitting looks fine, but the hyperfine splitting
is clearly too small (10.3 MeV).
We also note that the entirely new term $\propto \gamma_{\mu}
\gamma_{\nu} \gamma_{\rho}$ has a positive influence on the spectrum.
Without of it, both types of splitting decrease by about one fourth.

Finally we consider one more point, $\beta =5.5$, where $\eta_{c}$
corresponds to $m\simeq 1.1$.
Now the lattice is even coarser, $a =0.28$ fm.
The hyperfine splitting decreases to 7.4 MeV,
such that its slope for decreasing lattice spacing points up,
as it should. 
It is hardly justified to extrapolate the 
connection of these two points by a straight line down to the continuum
limit. If one does so nevertheless one obtains
$M_{J/\psi}-M_{\eta_{c}} \sim 28$ MeV.
The reference value here is perhaps not so much the experimental
value of $118 \pm 2$ MeV, but rather the relativistic quenched
results $\sim 70$ MeV; for a review see \cite{splitrev}.
NRQCD reported a larger value for some time, but more recent
observations indicate that the NR expansion is not really applicable 
to charmonium \cite{splitNR}.


To {\em summarize} our first experience with the hypercube
fermion plus truncated vertex function, we observed that
properties related to the restoration of Lorentz symmetry
are strongly improved: fermionic dispersion, kinetic mass
and mesonic dispersion. However, properties related to
the magnetic interaction are only successful in part:
$m_{B}$ is significantly improved, but the hyperfine splitting
is too small. The 2s-1s splitting looks fine. On the other hand,
the mass is still strongly renormalized.

Since renormalization due to $O(A^{2})$ is not incorporated 
yet, we expect tadpole improvement 
to amplify the hyperfine splitting and to reduce  
the dependence on the lattice spacing.
We also hope to further improve the parameterization.

But even if this action should not be satisfactory in a direct 
application, we expect it to be a good starting point for
the non-perturbative multigrid improvement program described
in \cite{LAT96}.
Furthermore it helped to identify the significant non-standard
terms, in particular the $c_{1}$ and the $s_{1}$ term.
Their dominating r\^{o}le in the improvement is a reliable
observation, even if the exact coefficients may still change.

We thank R. Edwards for allowing us to use the SZIN package
and his fitting routines. Much of the computation was
done at the Theoretical Physics Computing Facility at Brown University.

\end{document}